\documentclass[11pt,twoside]{article}
\usepackage{./asp2014}

\aspSuppressVolSlug
\resetcounters

\begin{document}

\title{ASTERICS -- Addressing Cross-Cutting Synergies and Common Challenges for the Next Decade Astronomy Facilities}
\author{Fabio Pasian,$^1$ Michael A. Garrett,$^2$ Francoise Genova,$^3$ Giovanni Lamanna,$^4$ Stephen Serjeant,$^5$ Arpad Szomoru,$^6$ and Rob van der Meer$^2$
\affil{$^1$INAF--OATS, Trieste, Italy; \email{pasian@oats.inaf.it}}
\affil{$^2$ASTRON, Dwingeloo, The Netherlands; \email{garrett@astron.nl}}
\affil{$^3$Observatoire astronomique de Strasbourg, Universit{\'e} de Strasbourg, CNRS, UMR 7550, Strasbourg, France; \email{francoise.genova@astro.unistra.fr}}
\affil{$^4$LAPP, Laboratoire d'Annecy-le-Vieux de Physique des Particules, Universit{\'e} Savoie Mont Blanc, CNRS/IN2P3, F-74941 Annecy-le-Vieux, France; \email{giovanni.lamanna@lapp.in2p3.fr}}
\affil{$^5$The Open University, Milton Keynes, United Kingdom; \email{stephen.serjeant@open.ac.uk}}
\affil{$^6$JIVE, Dwingeloo, The Netherlands; \email{szomoru@jive.eu}}}

\paperauthor{Fabio Pasian}{pasian@oats.inaf.it}{0000-0002-4869-3227}{INAF -- Osservatorio Astronomico di Trieste}{ }{Trieste}{ }{34143}{Italy}
\paperauthor{Michael A. Garrett}{garrett@astron.nl}{ORCID_Or_Blank}{ASTRON}{ }{Dwingeloo}{ }{7991PD}{The Netherlands}
\paperauthor{Francoise Genova}{francoise.genova@astro.unistra.fr}{0000-0002-6318-5028}{Observatoire astronomique de Strasbourg, Universit{\'e} de Strasbourg, CNRS, UMR 7550}{Observatoire Astronomique de Strasbourg}{Strasbourg}{N/A}{F-67000}{France}
\paperauthor{Giovanni Lamanna}{lamanna@lapp.in2p3.fr}{ORCID_Or_Blank}{LAPP}{ }{Annecy-le-Vieux}{ }{74941}{France}
\paperauthor{Stephen Serjeant}{stephen.serjeant@open.ac.uk}{ORCID_Or_Blank}{The Open University}{Department of Physical Sciences}{Milton Keynes}{ }{MK7 6AA}{United Kingdom}
\paperauthor{Arpad Szomoru}{szomoru@jive.eu}{ORCID_Or_Blank}{JIVE}{R\&D Department}{Dwingeloo}{ }{7991PD}{The Netherlands}
\paperauthor{Rob van der Meer}{meer@astron.nl}{ORCID_Or_Blank}{ASTRON}{ }{Dwingeloo}{ }{7991PD}{The Netherlands}

\begin{abstract}
The large infrastructure projects for the next decade will allow a new quantum leap in terms of new possible science. ESFRI, the European Strategy Forum on Research Infrastructures, a strategic initiative to develop the scientific integration of Europe, has identified four facilities (SKA, CTA, KM3Net and E-ELT) deserving priority in support. The ASTERICS project aims to address the cross-cutting synergies and common challenges shared by the various Astronomy ESFRI and other world-class facilities. The project (22 partners across Europe) is funded by the EU Horizon 2020 programme with 15 MEuro in 4 years. It brings together for the first time the astronomy, astrophysics and particle astrophysics communities, in addition to other related research infrastructures. 
\end{abstract}

\section{Objectives}
The European Strategy Forum on Research Infrastructures (ESFRI), a strategic initiative to develop the scientific integration of Europe, has identified four facilities  whose science cases are so outstanding that they can be considered as the main (ground-based) priorities of the European Astronomy and Astro-particles communities. These are the Square Kilometre Array (SKA), the Cherenkov Telescope Array (CTA), KM3NeT (Km$^3$ Neutrino Telescope) and the European Extremely Large Telescope (E-ELT). 

To address the common challenges of these infrastructures through synergy, the ASTERICS (Astronomy ESFRI and Research Infrastructure Cluster) project was proposed to the European Commission (EC) and funded. Its major objectives are to support and accelerate the implementation of the ESFRI telescopes, to enhance their performance beyond the current state-of-the-art, and to see them interoperate as an integrated, multi-wavelength and multi-messenger facility. An important focal point is the management, processing and scientific exploitation of the huge datasets the ESFRI facilities will generate. ASTERICS will seek solutions to these problems outside of the traditional channels by directly engaging and collaborating with industry and specialised small and medium-sized enterprises (SMEs). 

The various ESFRI pathfinders and precursors and other selected world-class projects (including space-borne facilities) will present the perfect proving ground for new methodologies and prototype systems. In addition, ASTERICS will enable astronomers from across the member states to have broad access to the reduced data products of the ESFRI telescopes via a seamless interface to the Virtual Observatory framework. This is expected to massively increase the scientific impact of the telescopes, and greatly encourage use (and re-use) of the data in new and novel ways, typically not foreseen in the original proposals. By demonstrating cross-facility synchronicity, and by harmonising various policy aspects, ASTERICS will realise a distributed and interoperable approach that ushers in a new multi-messenger era for astronomy. 

Through an active dissemination programme, including direct engagement with all relevant stakeholders, and via the development of citizen scientist mass participation experiments, ASTERICS has the ambition to be a flagship for the scientific, industrial and societal impact ESFRI projects can deliver.

\section{Project organisation}
The work packages have been built around the astronomy related ESFRI projects with a sharp focus on advancing and contributing to their design, construction and implementation. Futhermore, the work package activities and their deliverables have been selected to be of high impact and broadly relevant to as wide a range of the astronomy related ESFRI projects as possible, addressing common technical challenges and collective issues such as harmonisation, interoperability, exchange and cross-facility multi-wavelength/multi-messenger integration. 

{\bf ASTERICS Management Support Team (AMST)} \\
This work package will establish the ASTERICS Management Support Team (AMST), and will thus guarantee the smooth execution of all financial, administrative and reporting elements of the project. It will also permit the AMST to exercise central control and oversight of the scientific and technical progress of the project, as measured by secured milestones and the successful receipt of deliverables. A high-level Policy Forum (involving the ESFRI projects and other large astronomy research infrastructures) will also be established in order to coordinate and agree new models for joint time allocation, observing and data access/sharing, in addition to other more general policy matters of common interest. The culmination of ASTERICS will be an Integrating Event to  show-case the results of the project and their relevance to the ESFRI telescopes and all other relevant stakeholders.

{\bf Dissemination, Engagement and Citizen Science (DECS)} \\
The objective of DECS is to promote ASTERICS and the ESFRI astronomy facilities it aims to serve. In particular, it aims to open-up the ESFRI facilities to all relevant stakeholders and the widest possible audience, including the public:
i) Production of high quality branding and promotional outreach materials; more ambitiously, DECS also embraces the adoption of the principles of Science 2.0; 
ii) Create web-based interfaces that will open-up the astronomy related ESFRI facilities to the general public via a harmonised suite of Citizen Science mass participation experiments (MPEs) and online video material; 
iii) Attract young people to science by networking the ESFRI facilities via citizen science initiatives, and through coordinating open educational resources.

{\bf OBservatory E-environments Linked by Common challengeS (OBELICS)} \\
The aim of OBELICS is to enable interoperability and software re-use for the data generation, integration and analysis of the ESFRI and pathfinder facilities through the creation of an open innovation environment for establishing open standards and software libraries for multiwavelength/multi-messenger data. 
The specific objectives are:
i) Train researchers and data scientists in the ASTERICS-related projects to apply state-of-the-art parallel software programming techniques, to adopt big-data software frameworks, to benefit from new processor architectures and e-science infrastructures. This will create a community of experts that can contribute across facilities and domains; 
ii) Maximise software re-use and co-development of technology for the robust and flexible handling of the huge data streams generated by the ASTERICS-related facilities (this involves the definition of open standards and design patterns, and the development of software libraries in an open innovation environment); 
iii) Adapt and optimise extremely large database systems to fulfil the requirements of the ASTERICS-related projects (this requires the development of use cases, prototypes and benchmarks to demonstrate scalability and deployment on distributed non-homogeneous resources); cooperation with the ESFRI pathfinders, computing centres, e-infrastructure providers and industry will be organised and managed to fulfil this objective; 
iv) Study and demonstrate data integration across ASTERICS-related projects using data mining tools and statistical analysis techniques on Petascale data sets (this will require adaptable and evolving workflow management systems, to allow deployment on existing and future e-science infrastructures).
All tasks are built upon the state-of-the-art in ICT, in cooperation with major European e-infrastructures and are conceived to minimise fragmentation. Communications and links with other communities and e-science service providers are considered in order to contribute to the effectiveness of the proposed objectives.

{\bf Data Access, Discovery and Interoperability (DADI)} \\
The Virtual Observatory (VO) framework is a key element in successfully clustering the ESFRI projects. With the ESFRI facilities and their pathfinders included in the VO, astronomers will be able to discover, access, use and compare their data, combining it with data from other ground- and space-based observatories as well as theoretical model collections. The goal of DADI is to make the ESFRI and pathfinder project data available for discovery and usage in the international VO framework, and accessible with the set of VO-enabled common tools. More specifically:
i) Train and support ESFRI projects staff in the usage and implementation of the VO framework and tools, and make them active participants in the development of the VO framework definition and updates, thus contributing to relevance and sustainability of the framework; 
ii) Train and support the wider astronomical community in scientific use of the framework, in particular for pathfinder data, and gather their requirements and feedback; 
iii) Adapt the VO framework and tools to the ESFRI projects needs, in continuous cooperation with the International Virtual Observatory Alliance (IVOA).

{\bf Connecting Locations of ESFRI Observatories and Partners in Astronomy for Timing and Real-time Alerts (CLEOPATRA)} \\
The partners in ASTERICS share an ambition to use modern communication methods, such as fast broadband connectivity, to improve the scientific capabilities of their research infrastructures. The research activities aim specifically at synergetic observing modes, and fast and reliable access to large data streams. These aspects are covered in CLEOPATRA: 
i) Develop technology for the enabling of long-haul and many-element time and frequency distribution over fibre connections. This has the potential to increase the efficiency and affordability of all radio astronomy facilities (SKA, LOFAR, EVN); such developments are also highly relevant for astroparticle facilities (CTA, KM3NET) and can enable novel realtime multi messenger observations; 
ii) Develop methods for relaying alerts, which will signal transient event detections between the facilities and enable joint observing programmes; the focus is both on interchange formats and on scientific strategies and methods for joint observing; 
iii) Further development of existing data streaming software, building on previous e-VLBI projects, and providing tools for robust and efficient data dissemination for all facilities in the user domain, including ESO facilities such as ALMA and the E-ELT; 
iv) Foster the development of advanced scheduling algorithms, using AI approaches for optimal usage of the ESFRI facilities, so to achieve a consistent set of enhancements of the facilities based on developments in connectivity and data transport.

{\bf Interactions among work packages} \\
OBELICS and DADI have a strong and interrelated focus on delivering common solutions, standards and analysis to the management and exploitation of large volume and high velocity data streams. Key goals include interoperability between facilities such as the SKA, CTA, Euclid, EGO, KM3Net etc. 
Both OBELICS and DADI also entail significant training opportunities for external stakeholders, in order to ensure that ESFRI projects staff and users are fully engaged with the ASTERICS programme.There is a direct connection between these activities, and a specific task in OBELICS is dedicated to the interface and coordination with DADI. 
Quite naturally, as all work packages need a dissemination activity to be carried out, and need to be managed, all of them will have a direct  connection to DECS and AMST. 

\section{Conclusions}
Using an inclusive, engaging and open approach in the preparation of the project work plan, the ASTERICS partners have arrived at a concept that is relevant to the implementation of the astronomy related ESFRI facilities, and that is supported and endorsed by the main players of the multi-disciplinary astronomy and astroparticle physics communities.

ASTERICS is an ambitious project but the goals are clear and attainable. The ultimate measure of its success is how often the project results and products are incorporated into the ESFRI observatories. On timescales similar to the duration of the project (4 years), success can also be measured via implementation on the ESFRI pathfinder telescopes -- the latter provide existing platforms on which ASTERICS technologies can be tested and proven. The incorporation of other related projects (e.g. Euclid, EGO) provides further testing opportunities, and additionally addresses the demand to realise the potential offered by combining astrophysics and cosmology together.

\acknowledgements ASTERICS is a project funded by the European Commission under the Horizon2020 programme (id 653477). All the partner institutions and individual participants are gratefully acknowledged for their work in the project.  



\end{document}